\documentclass[conference]{IEEEtran}
\usepackage[dvips]{graphicx}
\usepackage{subfigure}
\usepackage{times}
\usepackage{latexsym}
\usepackage{amsmath}
\usepackage{amssymb}
\usepackage{multirow}
\usepackage{algorithmic}
\usepackage{eucal}
\usepackage{epsfig}
\usepackage{cite}
\usepackage{longtable}
\usepackage{slashbox}
\usepackage{amsthm}
\usepackage{latexsym,bm}
\usepackage{color}
\usepackage{subfigure}
\usepackage{multicol}
\usepackage{slashbox}

\theoremstyle{plain}

\theoremstyle{definition}
\theoremstyle{definition}

   \def\bu{{\mathbf{u}}}
\def\bb{{\mathbf{b}}}  \def\bp{{\mathbf{p}}} \def\bv{{\mathbf{v}}}
\def\bc{{\mathbf{c}}}  \def\bq{{\mathbf{q}}} \def\bw{{\mathbf{w}}}
  \def\br{{\mathbf{r}}} \def\bx{{\mathbf{x}}}
  \def\bs{{\mathbf{s}}} 
   
\def\bg{{\mathbf{g}}} \def\bn{{\mathbf{n}}}

 \def\bH{{\mathbf{H}}}  
  \def\bP{{\mathbf{P}}}

\begin{document}

\title{Performance of Joint Channel and Physical Network Coding Based on Alamouti STBC}
\author{\IEEEauthorblockN{Yi Fang and Lin Wang}
\IEEEauthorblockA{Department of communication Engineering\\
Xiamen University, Xiamen, China\\
Email: wanglin@xmu.edu.cn, fangyi1986812@gmail.com}
\and
\IEEEauthorblockN{Kai-Kit Wong and Kin-Fai Tong}
\IEEEauthorblockA{Department of Electronic and Electrical Engineering\\
University College London, London, UK\\
Email:  kai-kit.wong@ucl.ac.uk, k.tong@ucl.ac.uk}}

\maketitle
\begin{abstract}
This work considers the protograph-coded physical network coding (PNC) based on Alamouti space-time block coding (STBC) over Nakagami-fading two-way relay channels, in which both the two sources and relay possess two antennas. We first propose a novel precoding scheme at the two sources so as to implement the iterative decoder efficiently at the relay. We further address a simplified updating rule of the log-likelihood-ratio (LLR) in such a decoder. Based on the simplified LLR-updating rule and Gaussian approximation, we analyze the theoretical bit-error-rate (BER) of the system, which is shown to be consistent with the decoding thresholds and simulated results. Moreover, the theoretical analysis has lower computational complexity than the protograph extrinsic information transfer (PEXIT) algorithm. Consequently, the analysis not only provides a simple way to evaluate the error performance but also facilitates the design of the joint channel-and-PNC (JCNC) in wireless communication scenarios.
\end{abstract}

\section{Introduction}
It is well known that physical network coding (PNC) can significantly enhance the throughput in networks compared to the conventional network coding scheme \cite{Zhang2006}. In PNC, the relay can employ the interference as useful information and hence decode the network-coded information (XOR information) rather than decode the separated information of the two sources. Inspired by such advantages, the PNC has been introduced for many wireless communication systems, e.g., ultra-wideband (UWB) communication systems \cite{6133672}.

To obtain more performance enhancement, a wealth of research has investigated on the interplay between PNC and other spatial diversity techniques \cite{6364770,5545660,6380933}, such as multiple-input multiple-output (MIMO) and space-time block coding (STBC). In particular, \cite{6364770} and \cite{6380933} have proposed the precoding scheme respectively for the MIMO-PNC system and the STBC-PNC system before transmission to facilitate the PNC operation at the relay.

In recent years, joint channel-and-PNC (JCNC) schemes have been carefully studied for different channel conditions, e.g. non-fading \cite{5072363,5493997} and fading scenarios \cite{5683819,6381454}. Assuming that the same channel code is employed by the two sources, the XOR codeword at the relay can be decoded using the modified iterative decoding algorithm \cite{5072363,5493997,5683819,6381454}. To further perform the analysis on such a decoder in the additive white Gaussian noise (AWGN) channels, \cite{6125282} has proposed a new method to model the extrinsic information transfer (EXIT) chart of JCNC.

However, to the best of our knowledge, there is little research in the literature considering the performance of the STBC-based JCNC (i.e., STBC-JCNC) systems. In this paper, we provide insightful study of a STBC-JCNC system in fading scenarios, where the protograph code is used as the channel code. It was pointed out that fading statistics in wireless communications are more suitably to be described by a Nakagami distribution \cite{380145}. Specially, the distribution of the small-scale amplitudes of IEEE 802.15.4a UWB channel models \cite{Molisch04ieee802.15.4a} is modeled as Nakagami. Based on the aforementioned observation, the presently proposed system will be investigated over fading channels subjected to Nakagami distribution. Assuming that the reciprocal channel state information (CSI) is available at the two sources, we design a novel precoding scheme at the sources such that the relay can decode the XOR codeword more efficiently. We also propose a simplified updating rule of the log-likelihood-ratio (LLR) message in the JCNC decoder. Furthermore, we derive the bit-error-rate (BER) formula utilizing the updating rule and Gaussian approximation. It is shown that the the BER formula is in agreement with  the simulated results and reduces the computational complexity as compared to the protograph extrinsic information transfer (PEXIT) algorithm \cite{Fang2012}.

This paper is organized as follows. In Section \ref{sect:system}, the proposed STBC-JCNC scheme is described. In Section \ref{sect:Performance}, we provide the simplified LLR-updating rule and analyze the system performance. Numerical results are carried out in Section \ref{sect:SIMU}, and conclusions are given in Section \ref{sect:Conclusions}.

\section{Description of the Proposed STBC-JCNC System}
\label{sect:system}

\subsection{System Model} \label{sect:model}
\begin{figure}[t]
\center
\includegraphics[width=3.0in,height=1.3in]{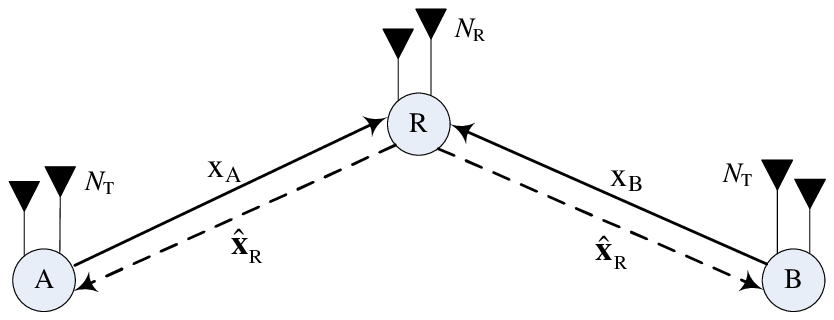}
\vspace{-0.3cm}
\caption{The system model of proposed STBC-JCNC system over the Nakagami fading channels. The MAC stage and the BC stage are denoted by the solid lines and the dotted lines, respectively.}
\label{fig:Fig.1}
\end{figure}

Consider an Alamouti STBC-based \cite{730453} two-way relay system with two sources (A and B) and one relay (R) in Fig.~\ref{fig:Fig.1}, in which all the terminals are equipped with
two antennas (i.e., $N_{\rm T}= N_{\rm R} =2$). For this system, both two sources are to obtain information from each other with the assistance of the relay. Let ${\bb}_{\rm Z} \in \{0,1\}^K$ (${\rm Z={A, B}}$) denotes the binary information vector of Z. We assume that the two sources adopt the same channel coding scheme to form the codewords with length $N$,
i.e., ${\bb}_{\rm Z} \to {\bc}_{\rm Z}$ ($\bc_{\rm Z} \in \{0,1\}^N$). Afterwards, the binary coded bits $c_{{\rm Z},j}$ ($j=1,2,\cdots$) are converted into $x_{{\rm Z},j} = (-1)^{c_{{\rm Z},j}}$ by a BPSK modulator and then mapped onto an $N_{\rm T} \times T$ STBC transmit
matrix ${\bx}_{\rm Z}$ ($T=2$ is the transmit time slot), denoted as $\bx_{\rm Z} =
\left[ \begin{array}{lr}
x_{{\rm Z},1} & -x_{{\rm Z},2}^{\ast} \cr
x_{{\rm Z},2} & x_{{\rm Z},1}^{\ast}
\end{array} \right]$ (where ``$^\ast$'' is the conjugate of a variable). As shown in \cite{6380933}, it is impossible to directly decode the XOR codeword
${\bc}_{{\rm A} \oplus {\rm B}}={\bc}_{\rm A}\oplus {\bc}_{\rm B}$ from the received signals at the relay. Moreover, to separately decode ${\bc}_{\rm A}$ and ${\bc}_{\rm B}$, the maximum-likelihood (ML) decoding \cite{5545660} is not applicable for JCNC, because ML decoding of general linear codes is well-known to be non-deterministic polynomial-time hard (NP-hard). To resolve this problem, the precoders for the STBC signals of the two sources should be properly designed before transmission in order to facilitate the decoding of ${\bc}_{{\rm A} \oplus {\rm B}}$. We first denote ${\bP}_{\rm Z}$ as the $N_{\rm T}\times N_{\rm T}$ precoding matrix for Z and will introduce the detailed design of the precoding matrices later.

The transmission scheme includes two stages: multiple access (MAC) and broadcast (BC). In the MAC stage, the two sources (i.e., A and B) transmit STBC signals to the relay simultaneously over Nakagami fading channels. For the $k$-th ($k=1, 2$) antenna at the relay, the receive signal vector $\br_{\rm R}[k]=[r_{{\rm R},1}[k]\;r_{{\rm R},2}[k]]$ of dimension $1\times T$ can be expressed as
\begin{equation}
\br_{\rm R}[k] = {\bH}_{\rm A}[k] {\bP}_{\rm A} {\bx}_{\rm A} + {\bH}_{\rm B}[k] {\bP}_{\rm B}  {\bx}_{\rm B} +{\bn}_{\rm R}[k]
\label{eq:receiver-r}
\end{equation}
where ${\bH}_{\rm Z}[k]= [h_{{\rm Z},1}[k]\;h_{{\rm Z},2}[k]]$ of dimension
$1\times N_{\rm T}$ is the channel fading vector, and
${\bn}_{\rm R}[k]= [n_{{\rm R},1}[k]\;h_{{\rm R},2}[k]]$ of dimension $1\times T$
is the AWGN noise vector whose elements follow the independent and identically
Gaussian distribution with zero mean and variance $\sigma_n^2$ ($\sigma_n^2=N_{0}/2$),
i.e., ${\cal N}(0,\sigma_n^2)$. Based on $\br_{\rm R}[k]$, the relay can use the
modified belief propagation (BP) algorithm \cite{5683819,6381454} to estimate the
XOR codeword ${\bc}_{{\rm A}\oplus {\rm B}}$,
denoted as ${\hat{\bc}}_{\rm R}={\hat {\bc}}_{{\rm A}\oplus {\rm B}}$. In the BC stage, the estimated information is re-encoded, modulated, and STBC-mapped. After that, the STBC signal $\hat{\bx}_{\rm R}$ is broadcasted to both the two sources. In this paper, we only focus on the MAC stage, because the decoding performance at the relay is critical to the whole
performance of such transmission mechanism \cite{5493997,5683819}.

\subsection{Design of the Precoding Matrices}
\label{sect:Design-precoder}
Firstly, we assume that each source know the perfect reciprocal CSI in our system. The design of the two precoders at the sources should satisfy two conditions.
One is that the total transmission power of each source should be kept as $E_b$; secondly, inspired by the maximum-ratio-combined (MRC) signals of
the conventional STBC system \cite{730453} and the principle of PNC \cite{Zhang2006}, the relay should try to group the messages from A and B, i.e., $x_{{\rm A},j}$ and $x_{{\rm B},j}$ ($j=1,2$), together. This operation can be facilitated by defining the precoders at A and B as two diagonal matrices
using \cite[eq.~(6)]{6380933}
\begin{subequations}
\begin{eqnarray}
{\bP}_{\rm A} =
\left[
\begin{array}{lr}
\eta_{\rm A}[k] h_{\rm B,1}[k] & 0 \cr
0 & \eta_{\rm A}[k] h_{\rm B,2}[k]
 \end{array}
\right]
\label{eq:PA}
\\
{\bP}_{\rm B} =
\left[
\begin{array}{lr}
\eta_{\rm B}[k] h_{\rm A,1}[k] & 0 \cr
0 & \eta_{\rm B}[k] h_{\rm A,2}[k]
 \end{array}
\right]
\label{eq:PB}
\end{eqnarray}
\end{subequations}
where $\eta_{\rm Z}[k]$ is the is the normalization factor that ensure the transmission power of each source to be $E_b$.

To find the appropriate normalization factors, we can express the transmission power constraint by
\begin{equation}
{\bf trace} \{ {\bP}_{\rm A} {\bP}_{\rm A}^{H}\}= {\bf trace} \{ {\bP}_{\rm B} {\bP}_{\rm B}^{H} \} = E_b
\label{eq:power-constrain}
\end{equation}
Here, ${\bf trace}(\cdot)$ denotes the trace of a matrix and the superscript ``\textit{H}'' denotes the transpose conjugate of a matrix or vector.
Substituting \eqref{eq:PA} and \eqref{eq:PB} into \eqref{eq:power-constrain} yields
\begin{subequations}
\begin{align}
\eta_{\rm A}[k] & = \frac{1} {\sqrt{|h_{\rm B,1}[k]|^2 + |h_{\rm B,2}[k]|^2}}\\
\eta_{\rm B}[k] & = \frac{1} {\sqrt{|h_{\rm A,1}[k]|^2 + |h_{\rm A,2}[k]|^2}}
\end{align}
\end{subequations}

Using such precoding matrices (i.e., \eqref{eq:PA} and \eqref{eq:PB}), the relay can
group the messages from A and B as the following
\begin{subequations}
\begin{align}
\hspace{-1mm}r_{\rm R,1}[k] =& \displaystyle h_{\rm A,1}[k] h_{\rm B,1}[k] (\eta_{\rm A}[k] x_{\rm A,1} + \eta_{\rm B}[k] x_{\rm B,1} )\nonumber\\
&+ h_{\rm A,2}[k] h_{\rm B,2}[k] (\eta_{\rm A}[k] x_{\rm A,2} + \eta_{\rm B}[k] x_{\rm B,2} ) + n_{\rm R,1} [k]\nonumber\\
=&  \displaystyle h_1[k] (\eta_{\rm A}[k] x_{\rm A,1} + \eta_{\rm B}[k] x_{\rm B,1} ) \nonumber\\
&+ h_2[k] (\eta_{\rm A}[k] x_{\rm A,2} + \eta_{\rm B}[k] x_{\rm B,2} ) + n_{\rm R,1} [k]
\label{eq:r-R1k}\\
\hspace{-1mm} r_{\rm R,2}[k] =& \displaystyle -h_{\rm A,1}[k] h_{\rm B,1}[k] (\eta_{\rm A}[k] x_{\rm A,2}^{\ast} + \eta_{\rm B}[k] x_{\rm B,2}^{\ast} ) \nonumber\\
&+ h_{\rm A,2}[k] h_{\rm B,2}[k] (\eta_{\rm A}[k] x_{\rm A,1}^{\ast} + \eta_{\rm B}[k] x_{\rm B,1}^{\ast} ) + n_{\rm R,2} [k]\nonumber\\
=&  \displaystyle -h_1[k] (\eta_{\rm A}[k] x_{\rm A,2} + \eta_{\rm B}[k] x_{\rm B,2} )^{\ast} \nonumber\\
&+ h_2[k] (\eta_{\rm A}[k] x_{\rm A,1} + \eta_{\rm B}[k] x_{\rm B,1} )^{\ast} + n_{\rm R,2} [k]
\label{eq:r-R2k}
\end{align}
\end{subequations}
where $h_\mu[k]=h_{{\rm A},\mu}[k] h_{{\rm B},\mu}[k]\,(\mu=1,2)$, $\eta_{\rm A}[k] x_{\rm A,1}^{\ast} + \eta_{\rm B}[k] x_{\rm B,1}^{\ast}=(\eta_{\rm A}[k] x_{\rm A,1} + \eta_{\rm B}[k] x_{\rm B,1} )^{\ast}$, and $\eta_{\rm A}[k] x_{\rm A,2}^{\ast} + \eta_{\rm B}[k] x_{\rm B,2}^{\ast} = (\eta_{\rm A}[k] x_{\rm A,2} + \eta_{\rm B}[k] x_{\rm B,2} )^{\ast}$ since $\eta_{\rm Z}[k]$ is the real-valued variable.

Therefore, the output signals of the MRC combiner are given as
\begin{subequations}
\begin{align}
y_{\rm R,1} =& \displaystyle \sum_{k=1}^{N_{\rm R}} \left( \left| h_1 [k] \right|^2 + \left| h_2 [k] \right|^2 \right) \left( \eta_{\rm A}[k] x_{\rm A,1} + \eta_{\rm B}[k] x_{\rm B,1} \right) \nonumber\\
&+ \displaystyle \sum_{k=1}^{N_{\rm R}} \left( h_1^\ast [k] n_{\rm R,1} [k] + h_2 [k] n_{\rm R,2}^\ast [k] \right)\nonumber\\
=&  \displaystyle \xi_{\rm A,1} x_{\rm A,1} + \xi_{\rm B,1} x_{\rm B,1} + \varphi_{\rm R,1}
\label{eq:y-R1}
\end{align}
\begin{align}
y_{\rm R,2} =& \sum_{k=1}^{N_{\rm R}} \left( \left| h_1 [k] \right|^2 + \left| h_2 [k] \right|^2 \right) \left( \eta_{\rm A}[k] x_{\rm A,2} + \eta_{\rm B}[k] x_{\rm B,2} \right) \nonumber\\
&+ \sum_{k=1}^{N_{\rm R}} \left( h_2^\ast [k] n_{\rm R,1} [k] - h_1 [k] n_{\rm R,2}^\ast [k] \right)\nonumber\\
=&  \displaystyle \xi_{\rm A,2} x_{\rm A,2} + \xi_{\rm B,2} x_{\rm B,2} + \varphi_{\rm R,2}
\label{eq:y-R2}
\end{align}
\end{subequations}
where $\xi_{\rm A,1}=\xi_{\rm A,2}=\sum_{k=1}^{N_{\rm R}}
\left( \left| h_1 [k] \right|^2 + \left| h_2 [k] \right|^2 \right) \eta_{\rm A}[k]$, $\xi_{\rm B,1}=\xi_{\rm B,2}=\sum_{k=1}^{N_{\rm R}}
\left( \left| h_1 [k] \right|^2 + \left| h_2 [k] \right|^2 \right) \eta_{\rm B}[k]$, $\varphi_{\rm R,1}=\sum_{k=1}^{N_{\rm R}} \left( h_1^\ast [k] n_{\rm R,1} [k] + h_2 [k] n_{\rm R,2}^\ast [k] \right)$,
and $\varphi_{\rm R,2}=\sum_{k=1}^{N_{\rm R}} \left( h_2^\ast [k] n_{\rm R,1} [k] - h_1 [k]
n_{\rm R,2}^\ast [k] \right)$.

Hence, the generalized expression of the $j$-th variable node (VN) ($j=1,2,\cdots)$ is obtained exploiting \eqref{eq:y-R1} and \eqref{eq:y-R2}, as
\begin{equation}
y_{{\rm R},j} = \xi_{{\rm A},j} x_{{\rm A},j} + \xi_{{\rm B},j} x_{{\rm B},j} + \varphi_{{\rm R},j}
\label{eq:generalized-y}
\end{equation}
where
\begin{align}
& \hspace{-0.5mm}\xi_{{\rm Z},j} =
\left\{
\begin{array}{ll}
\displaystyle \sum_{k=1}^{N_{\rm R}} \left( \left| h_j [k] \right|^2 + \left| h_{j+1} [k] \right|^2 \right)\eta_{{\rm Z},j}[k],\, j = 2q-1
\\ \displaystyle \sum_{k=1}^{N_{\rm R}} \left( \left| h_{j-1} [k] \right|^2 + \left| h_j [k] \right|^2 \right)\eta_{{\rm Z},j}[k],\, j = 2q \nonumber
\end{array}
\right.
\\
& \hspace{-1.2mm} \varphi_{{\rm R},j} =
\left\{
\begin{array}{ll}
\displaystyle \sum_{k=1}^{N_{\rm R}} \left( h_j^\ast [k] n_{{\rm R},j} [k] + h_{j+1} [k] n_{{\rm R},j+1}^\ast [k] \right),\, j = 2q-1
\\ \displaystyle \sum_{k=1}^{N_{\rm R}} \left( h_j^\ast [k] n_{{\rm R},j-1} [k] - h_{j-1} [k] n_{{\rm R},j}^\ast [k] \right),\, j = 2q \nonumber
\end{array}
\right.
\end{align}
Here, $q=1,2,\cdots,\lceil\frac{N}{2}\rceil$ ($\lceil \frac{N}{2} \rceil$ denotes the smallest integer not less than $\frac{N}{2}$).
We can easily get ${\mathbb E} [\varphi_{{\rm R},j}]=0$ and ${\rm var}[\varphi_{{\rm R},j}] =\alpha_j \sigma_n^2$ ($\alpha_j$ is the short-hand notation for $\sum_{k=1}^{N_{\rm R}} \left( \left| h_j [k] \right|^2 + \left| h_{j+1} [k] \right|^2 \right))$. Note also that the index $j$ will be omitted in the remainder of the paper unless otherwise stated.

As seen from \eqref{eq:generalized-y}, there are four possible noise-free values for the received signal at the relay in such a system, i.e., ${\bs} = [s(0), s(1), s(2), s(3)]=[ \xi_{\rm A}+ \xi_{\rm B}, -\xi_{\rm A}+ \xi_{\rm B}, \xi_{\rm A}-\xi_{\rm B}, -\xi_{\rm A}- \xi_{\rm B}]$. Assuming that the input ({\it a-prior}) probability of $s(\mu) (\mu=0,1,2,3)$ is equal to $\frac{1} {4}$, we can compute the initial channel information of the $j$-th VN, i.e., $\bg = [g_0,g_1,g_2,g_3]$, as \cite{5683819}
\begin{align}
g_\mu = \beta e^{-\frac{(y_{\rm R}-s(\mu))^2} {2 \alpha \sigma_n^2}},~~ \mu=0,1,2,3
\label{eq:g-k}
\end{align}
where the normalization factor $\beta$ is used to ensure $\sum_{\mu=0}^{3}g_\mu=1$.

\section{Performance Analysis} \label{sect:Performance}

\subsection{Simplified LLR-Updating Rule} \label{sect:LLR-rule}
Similar to \cite{6125282}, we define one primary LLR, i.e., $L_{\rm P}$, and two secondary LLRs, i.e., $\rho_{{\rm P},1}$ and $\rho_{{\rm P},2}$, for a given (probability) message $\bP = [P_0, P_1, P_2, P_3]$ as follows, $L_{\rm P}={\rm In}(\frac{P_0+P_3}{P_1+P_2})$, $\rho_{{\rm P},1}={\rm In} (\frac{P_0}{P_3})$ and  $\rho_{{\rm P},2}={\rm In} (\frac{P_1}{P_2})$. To simplify the analysis, we assume that all the input secondary LLRs are kept $0$.\footnote{ This assumption will not affect the accuracy of our analytical result \cite{6125282}.} We also assume that all-zero (infinite) codeword is transmitted. In the following, we will discuss the simplified updating rule of the (primary) LLR in the JCNC decoder \cite{5683819} which is necessary for performing the performance analysis.
\\{\bf 1) Initialization:} The initial LLR of the $j$-th VN is computed based on \eqref{eq:g-k}
\begin{align}
L_{\rm g} =& {\rm In} \left[\frac{g_0 + g_3} {g_1 + g_2}\right]
= {\rm In} \left[ \frac{\beta e^{-\frac{(y_{\rm R}-s(0))^2} {2 \alpha \sigma_n^2}}+\beta e^{-\frac{(y_{\rm R}-s(3))^2} {2 \alpha \sigma_n^2}}} {\beta e^{-\frac{(y_{\rm R}-s(1))^2} {2 \alpha \sigma_n^2}} + \beta e^{-\frac{(y_{\rm R}-s(2))^2} {2 \alpha \sigma_n^2}} } \right] \nonumber\\
=& \frac{s^2(1)-s^2(0)} {2 \alpha  \sigma_n^2} + {\rm In} \left[ e^{\frac{s(0) y_{\rm R}} {\alpha \sigma_n^2}} + e^{-\frac{s(0) y_{\rm R}} {\alpha \sigma_n^2}} \right] \nonumber\\
&- {\rm In} \left[ e^{\frac{s(1) y_{\rm R}} {\alpha \sigma_n^2}} + e^{-\frac{s(1) y_{\rm R}} {\alpha \sigma_n^2}} \right]
\label{eq:L-ch-1}
\end{align}
In the high signal-to-noise ratio (SNR) region, \eqref{eq:L-ch-1} can be reduced to
\begin{align}
\hspace{-0.75mm}L_{\rm g} & \approx\hspace{-0.75mm}
\left\{ \begin{array}{ll}
\hspace{-1.5mm}\frac{s^2(1)-s^2(0)} {2 \alpha \sigma_n^2} + \frac{s(0) y_{\rm R}} { \alpha \sigma_n^2} - \frac{s(1) y_{\rm R}} { \alpha \sigma_n^2},\, s(0) y_{\rm R} \ge 0, s(1) y_{\rm R} \ge 0 \\
\hspace{-1.5mm}\frac{s^2(1)-s^2(0)} {2 \alpha \sigma_n^2} - \frac{s(0) y_{\rm R}} { \alpha \sigma_n^2} - \frac{s(1) y_{\rm R}} { \alpha \sigma_n^2},\, s(0) y_{\rm R} < 0, s(1) y_{\rm R} \ge 0 \\
\hspace{-1.5mm}\frac{s^2(1)-s^2(0)} {2 \alpha \sigma_n^2} + \frac{s(0) y_{\rm R}} { \alpha \sigma_n^2} + \frac{s(1) y_{\rm R}} { \alpha \sigma_n^2},\, s(0) y_{\rm R} \ge 0, s(1) y_{\rm R} < 0 \\
\hspace{-1.5mm}\frac{s^2(1)-s^2(0)} {2 \alpha \sigma_n^2} - \frac{s(0) y_{\rm R}} { \alpha \sigma_n^2} + \frac{s(1) y_{\rm R}} { \alpha \sigma_n^2},\, s(0) y_{\rm R} < 0, s(1) y_{\rm R} < 0 \nonumber
\end{array}
\right. \\
&= \frac{s^2(1)-s^2(0)} {2 \alpha \sigma_n^2} + \frac{|s(0) y_{\rm R}|} {\alpha \sigma_n^2} - \frac{|s(1) y_{\rm R}|} {\alpha  \sigma_n^2}
\label{eq:L-ch-2}
\end{align}

For a fixed channel realization (fixed $\xi_{\rm A}$ and $ \xi_{\rm B}$), substituting $x_{\rm A} = x_{\rm B} = +1$  into \eqref{eq:generalized-y} yields $y_{\rm R}=s(0)+ n_{\rm R}$. Afterwards, \eqref{eq:L-ch-2} can be written as
\begin{align}
 L_{\rm g} = \frac{s^2(1)-s^2(0)} {2 \alpha \sigma_n^2} + \frac{(|s(0)|-|s(1)|)|s(0) + n_{\rm R}|} {\alpha \sigma_n^2}
\label{eq:L-ch-simple}
\end{align}
where ${\mathbb E} [L_{\rm g}] = \frac{(|s(0)|-|s(1)|)^2}{2 \alpha \sigma_n^2}$ and ${\rm var} [L_{\rm g}] = \frac{(|s(0)|-|s(1)|)^2}{\alpha \sigma_n^2}$, i.e., $L_{\rm g}$ follows a symmetric Gaussian distribution. Note that $L_{\rm g}=0$ if the VN is punctured.

\begin{figure}[tbp]
\center
\includegraphics[width=3.5in,height=2.5in]{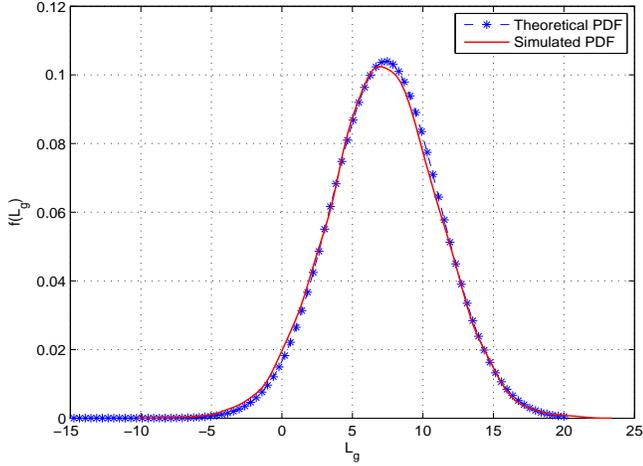}
\vspace{-0.4cm}
\caption{PDFs of $L_{\rm g}$ and ${\cal N}(u_{\rm g}, 2u_{\rm g})$ in the proposed STBC-JCNC system ($N_{\rm R}=2$) over Nakagami fading channel. The parameters used are  $m=2$ and $E_b/N_0=1.0$~dB.}
\label{fig:Fig.2}
\end{figure}

Using the rate-$0.8$ (all-zero) accumulate-repeat-by-3-accumulate (AR3A) protograph code
\cite{4155107} with information length of $K=6400$, we further validate symmetric Gaussian
distribution of the $L_{\rm g}$ values exploiting Monte Carlo
simulations and show the results in Fig.~\ref{fig:Fig.2}. A Nakagami fading channel with parameters  $m=2$ and $E_b/N_0=1.0$~dB is assumed.
We first plot the the simulated probability density function (PDF) of $L_{\rm g}$ (denoted by $f(L_{\rm g})$). Moreover, we also define $u_{\rm g} = {\mathbb E} [L_{\rm g}]$ and plot the PDF of the symmetric Gaussian distribution
${\cal N}(u_{\rm g}, 2u_{\rm g})$ in the same figure for comparison.
The curves in the figure indicate that the PDF of $L_{\rm g}$ agrees well with the PDF of
${\cal N}(u_{\rm g}, 2u_{\rm g})$, which suggests that $L_{\rm g}$ approximately follows a symmetric Gaussian distribution. Simulations have also been performed for other lower $E_b/N_0$ values and similar observations are obtained.
Accordingly, we will use the symmetric Gaussian assumption in the following analysis.
\\{\bf 2) LLR out of check nodes (CNs):} Consider a CN with degree $d_c=3$. Let $\bp = [p_0, p_1, p_2, p_3]$ and $\bq = [q_0, q_1, q_2, q_3]$ denote the input messages from two VNs. The output message from this CN to the $3$-rd VN, i.e., $\bu= [u_0, u_1, u_2, u_3]$, is then given by \cite[eq.~(17)]{5683819}
\begin{align}
\bu =&[p_0 q_0 + p_1 q_1 + p_2 q_2 + p_3 q_3, p_0 q_1 + p_1 q_0 + p_2 q_3 + p_3 q_2,\nonumber\\
& \hspace{0.5mm}p_0 q_2 + p_1 q_3 + p_2 q_0 + p_3 q_1, p_0 q_3 + p_1 q_2 + p_2 q_1 + p_3 q_0]\nonumber
\end{align}
The corresponding LLR of $\bu$ is calculated as
\begin{align}
L_{\rm o,u} = & {\rm In} \left[\frac{u_0 + u_3} {u_1 + u_2}\right]\nonumber\\
=& {\rm In} \left[\frac{(p_0 + p_3) (q_0 + q_3) + (p_1 + p_2) (q_1 + q_2) } {(p_0 + p_3) (q_1 + q_2) + (p_1 + p_2) (q_0 + q_3)} \right]\nonumber\\
=&{\rm In} \left[\frac{1 + e^{L_{\rm i,p} + L_{\rm i,q}}} {e^{L_{\rm i,p}} + e^{L_{\rm i,q}}} \right] = L_{\rm i,p} \boxplus L_{\rm i,q}
\label{eq:L-u}
\end{align}
where the subscripts ``i'' and ``o'' denote the ``input'' LLR and ``output'' LLR, respectively; and the boxplus operator ``$\boxplus$'' is defined in \cite{485714}. For a CN with degree $d_c>3$, the output LLR from this CN to the $d_c$-th VN is given by
\begin{align}
L_{\rm o, u} = \underbrace{L_{\rm i,p} \boxplus L_{\rm i,q} \boxplus\cdots \boxplus L_{\rm i,r}}_{d_c - 1\,{\rm terms}}
\label{eq:L-u-generalize}
\end{align}
\\{\bf 3) LLR out of VNs:} Suppose the $j$-th VN with degree $d_v=2$. Let $\bg = [g_0, g_1, g_2, g_3]$ and $\bw = [w_0, w_1, w_2, w_3]$  represent its channel message and the input message from one CN, respectively. Exploiting \cite[eq.~(14)]{5683819}, the output LLR from this VN to the other CN (the corresponding message is: $\bv =[v_0, v_1, v_2, v_3]$) is formulated as
\begin{align}
L_{\rm o,v} &= {\rm In} \left[\frac{v_0 + v_3} {v_1 + v_2}\right] = {\rm In} \left[\frac{g_0 w_0 + g_3 w_3} {g_1 w_1 + g_2 w_2}\right] \nonumber\\
&= L_{\rm g} + L_{\rm i,w} + K_{\rm v}
\label{eq:L-v}
\end{align}
where
\begin{align}
\hspace{-2mm} K_{\rm v} \hspace{-1mm}= \hspace{-1mm}{\rm In} \hspace{-0.5mm} \left[\frac{1 + e^{\rho_{\rm i,g,1} + \rho_{\rm i,w,1}}} {(1 + e^{\rho_{\rm i,g,1}})(1 + e^{\rho_{\rm i,w,1}})} \right] \hspace{-1mm}-\hspace{-1mm} {\rm In}\hspace{-0.5mm} \left[\frac{1 + e^{\rho_{\rm i,g,2} + \rho_{\rm i,w,2}}} {(1 + e^{\rho_{\rm i,g,2}})(1 + e^{\rho_{\rm i,w,2}})} \right]\nonumber
\label{eq:L-v}
\end{align}
Since the input secondary LLRs $\rho_{\rm i,g,1}$, $\rho_{\rm i,g,2}$, $\rho_{\rm i,w,1}$, $\rho_{\rm i,w,2}$ are assumed to be $0$, we have $ K_{\rm v}=0$ and hence $L_{\rm o, v} \approx L_{\rm g} + L_{\rm i,w}$. For a VN with degree $d_v>2$, the output LLR from this VN to the $d_v$-th CN is expressed as
\begin{equation}
L_{\rm o, v} \approx \underbrace{L_{\rm g} + L_{\rm i, w} +\cdots + L_{\rm i, t}}_{d_v\,{\rm terms}}
\label{eq:L-v-generalize}
\end{equation}
Note that the \textit{a-posteriori} LLR of the $j$-th VN is defined as $L_{\rm app, v} = L_{\rm o, v} +L_{\rm i, v}$. With the help of (\ref{eq:L-u-generalize}), (\ref{eq:L-v-generalize}), and the Gaussian distribution of the initial LLR, we demonstrate that the simplified updating rule of the LLR for the JCNC is the same as that for channel coding \cite{485714}. Consequently, we can apply the PEXIT algorithm \cite{Fang2012} for calculating the decoding threshold of the STBC-based protograph-coded PNC scheme over Nakagami fading channels.

\subsection{BER Expression} \label{sect:BER-analysis}
For a LDPC code corresponding to a protograph with $N$ VNs \cite{4155107}, assuming that  $\bar{L}_{{\rm app},j}$ is the expected \textit{a-posteriori} LLR of the $j$-th VN, and $\bar{I}_{{\rm app},j}$ is the expected \textit{a-posteriori} mutual information (MI) between $\bar{L}_{{\rm app},j}$ and the corresponding coded bit.\footnote{$\bar{L}_{{\rm app},j}= \frac{1} {Q} \sum_{q=1}^{Q} L_{{\rm app},j,q}$ ($Q$ is the total number of channel realizations \cite{Fang2012}) for the ergodic (i.e., fast-fading) scenario while $\bar{L}_{{\rm app},j}= L_{{\rm app},j}$ for the quasi-static case. Similar definition is used for $\bar{I}_{{\rm app},j}$.} Then, $\bar{L}_{{\rm app},j}$ can be calculated using the symmetric Gaussian assumption, resulting in \cite{Fang2013}
\begin{align}
{\rm var} [\bar{L}_{{\rm app},j}] =  \left\{ J^{-1} (\bar{I}_{{\rm app},j}) \right\}^2
\end{align}
where the function $J^{-1}(\cdot)$ and $\bar{I}_{{\rm app},j}^{t+1}$ are given in \cite[eq.~(4)]{6253209} and \cite[eq.~(22)]{Fang2012}, respectively.
\\We further evaluate the BER of the $j$-th VN after $t$ iterations as
\begin{align}
P_{b,j}^{t+1}
=& \frac{1} {2} \text{erfc} \left(\frac{{\mathbb E}[\bar{L}_{{\rm app},j}^{t+1}|c_{{{\rm A}\oplus {\rm B}},j}=0]} {\sqrt{2 {\rm var}[\bar{L}_{{\rm app},j}^{t+1}|c_{{{\rm A}\oplus {\rm B}},j}=0]}} \right) \nonumber\\
=& \frac{1} {2} \text{erfc} \left(\frac{\sqrt{{\rm var}[\bar{L}_{{\rm app},j}^{t+1}|c_{{{\rm A}\oplus {\rm B}},j}=0]}} {2 \sqrt{2}} \right)\nonumber\\
=& \frac{1} {2} \text{erfc} \left(\frac{\sqrt{{\rm var}[\bar{L}_{{\rm app},j}^{t+1}]}} {2 \sqrt{2}} \right) = \frac{1} {2} \text{erfc} \left(\frac{ J^{-1} (\bar{I}_{{\rm app},j}^{t+1})} {2 \sqrt{2}} \right)
\label{eq:Pb-bit}
\end{align}
where $\text{erfc}(\cdot)$ is the complementary error function, and ${\rm var}[\bar{L}_{{\rm app},j}^{t+1}|c_{{{\rm A}\oplus {\rm B}},j}]={\rm var}[\bar{L}_{{\rm app},j}^{t+1}]$ due to the all-zero codeword assumption.
\\Finally, the averaged BER over all $N$ VNs after $t$ iterations is written as
\begin{align}
\bar{P}_b^{t+1} = \frac{1} {N} \sum_{j=1}^{N} P_{b,j}^{t+1}
=  \frac{1} {2 N} \sum_{j=1}^{N}  \text{erfc} \left(\frac{ J^{-1} (\bar{I}_{{\rm app},j}^{t+1})} {2 \sqrt{2}} \right)
\label{eq:Pb-average}
\end{align}
{\bf Note also the following}

The maximum number of iterations of the theoretical BER analysis $T_{\rm max}$ ($t=T_{\rm max}$) equals to that of the simulations while the maximum number of iterations of the PEXIT algorithm $T_{\rm max}^{\rm p}$ should be much larger.\footnote{According to \cite{Fang2012}, the maximum iteration number of the PEXIT algorithm  $T_{\rm max}^{\rm p}$ should be large enough in order to ensure the complete convergence of the decoder, i.e., $T_{\rm max}^{\rm p} \ge 500$.} In this paper, we set $T_{\rm max} =100$ as in \cite{Fang2012,5751586}. Thus, the BER analysis is able to reduce computational complexity  by $\frac{T_{\rm max}^{\rm p}-T_{\rm max}} {T_{\rm max}^{\rm p}}$ as compared to the modified PEXIT algorithm. Moreover, the BER analysis can be used to evaluate the error performance for any $E_b/N_0$ and $T_{\rm max}$ rather than provide an $E_b/N_0$ threshold for sucessfully decoding.

In summary, the BER analysis is a more generalized analytical tool with lower
computational complexity.

\section{Numerical Results} \label{sect:SIMU}
Here, we firstly analyze the convergence performance of three different  JCNC schemes, i.e., single-input single-output JCNC (SISO-JCNC) \cite{5683819}, STBC-JCNC with $N_{\rm R}=1$, the proposed STBC-JCNC ($N_{\rm R}=2$) utilizing the PEXIT algorithm \cite{Fang2012}. Then, we perform simulations  to compare the BER performance of the STBC-JCNC systems with the SISO-JCNC system and to validate the theoretical analyses in Sect.~\ref{sect:Performance}. In all the numerical results, we use the rate-$0.8$ AR3A code \cite{4155107} as the channel code. The channel being considered is the ergodic Nakagami-fading two-way relay channel.\footnote{Although we consider the ergodic scenario in simulations, the analytical result in this paper is also applicable for the quasi-static case (where the fading parameter of each link is kept constant for a code block).}

\subsection{Convergence Performance}
We investigate the convergence performance of the proposed STBC-JCNC scheme in terms of the decoding threshold. It is shown that low threshold indicates relatively good convergence performance \cite{957394}. As a comparison, we also consider the conventional SISO-JCNC scheme \cite{5683819} and the STBC-JCNC scheme with $N_{\rm R}=1$. The results are shown in Table~\ref{tab:table1}. Given a fixed fading depth (fixed $m$), we can observe that the schemes, from the best to the worst convergence performance, are in the following order: (i) proposed STBC-JCNC, (ii) STBC-JCNC with $N_{\rm R}=1$, and (iii) SISO-JCNC.
\begin{table}[t]
\caption{Decoding thresholds $(E_b/N_0)_{\rm th}~({\rm dB})$ of three different JCNC schemes over Nakagami fading channels. The rate-$0.8$ AR3A code is used.}
\begin{center}
\begin{tabular}{|c|c|c|c|c|}
\hline
\backslashbox{JCNC-scheme}{$m$} & $1$ &  $2$  & $3$  \\\hline
SISO-JCNC & $9.125$ & $5.982$ & $4.894$ \\\hline
STBC-JCNC with $N_{\rm R}=1$ & $6.465$ & $4.528$ & $3.915$ \\\hline
Proposed STBC-JCNC & $1.893$ & $0.771$ & $0.381$ \\\hline
\end{tabular}
\end{center}
\label{tab:table1}
\end{table}%

\subsection{BER Performance}
In the sequel, we perform simulations of SISO-STBC and STBC-JCNC schemes to
test their BER performance. Further, we compare the simulated results with the theoretical
BER performance and the decoding thresholds to verify the accuracy of our analysis. The
rate-$0.8$ AR3A code  with parameters $[N, K, P]=[8800, 6400, 800]$ ($P$ is the punctured length) is used as the channel code. Unless specified otherwise, it has been assumed that the decoder
performs a maximum of $100$ iterations for each code block.

\begin{figure}[tbp]
\center
\includegraphics[width=3.5in,height=2.5in]{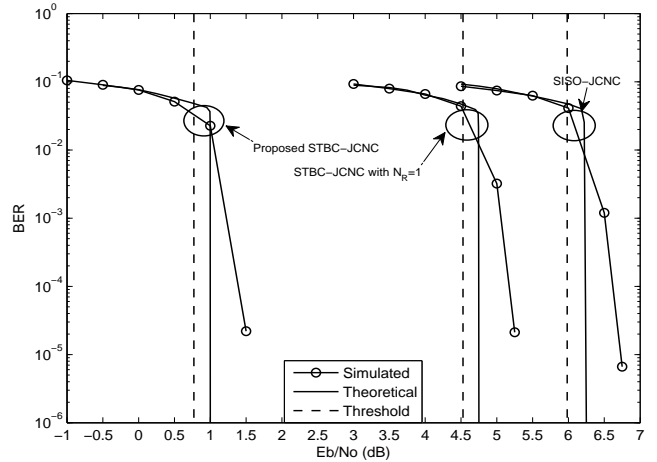}
\vspace{-0.3cm}
\caption{BER performance of SISO-JCNC and the STBC-JCNC systems over Nakagami fading channels with the fading depth $m=2$.
}
\label{fig:Fig.3}
\end{figure}

Fig.~\ref{fig:Fig.3} presents the BER performance of the SISO-JCNC and the STBC-JCNC
systems. Referring to this figure, the proposed STBC-JCNC system
is the best-performing scheme among the three schemes. For example, to achieve a
BER of $2 \times 10^{-5}$, the required $E_b/N_0$ of the SISO-JCNC, STBC-JCNC with
$N_{\rm R}=1$, and proposed STBC-JCNC are respectively $6.65$~dB, $5.25$~dB, and $1.5$~dB. In the same figure, at a BER of $2 \times 10^{-5}$, the threshold
agrees well with the corresponding analytical $E_b/N_0$ within $0.3$~dB for all the BER curves.
Furthermore, the difference between the simulated $E_b/N_0$ and the analytical $E_b/N_0$
is around $0.4-0.5$~dB, which is shown to be consistent with the results in
\cite{5751586}. It is because the finite-block-length codeword is used for the simulations.

\begin{figure}[tbp]
\center
\includegraphics[width=3.5in,height=2.5in]{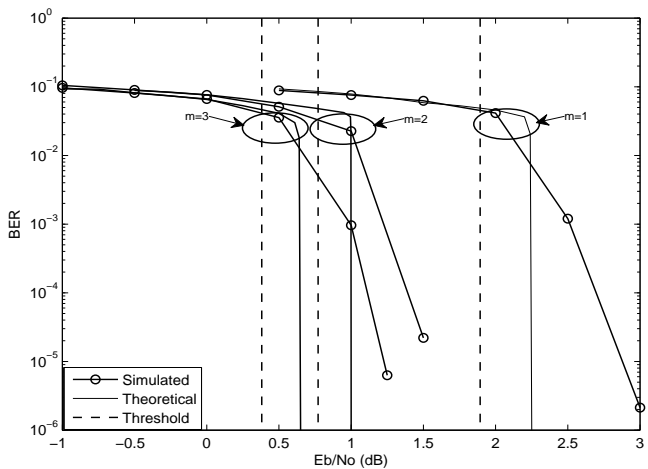}
\vspace{-0.3cm}
\caption{BER performance of the proposed STBC-JCNC system over Nakagami fading channels with different fading depths.
}
\label{fig:Fig.4}
\end{figure}

Fig.~\ref{fig:Fig.4} further compares the decoding thresholds, the theoretical and
simulated BER results of the proposed STBC-JCNC system over Nakagami
fading channels with different fading depths (different values of $m$). As seen from
this figure, for all the fading depths, at a BER of $2 \times 10^{-5}$, the threshold,
the theoretical $E_b/N_0$, and the simulated $E_b/N_0$ are reasonably
consistent. Moreover, the the performance of the system is improved as the fading depth
decreases (higher $m$), but the increment of the gain is reduced as $m$ becomes larger.
These observations agree well with the results of threshold analysis in
Table~\ref{tab:table1}.

\section{Conclusions} \label{sect:Conclusions}
In this paper, we have insightfully investigated the STBC-JCNC system over Nakagami fading channels. We have first proposed a novel precoder for the system such that the relay can easily decode the PNC codeword. Subsequently, we have developed a simplified updating rule of the LLR in the JCNC iterative decoder and hence derive the BER formula of the proposed system. Results have shown that the theoretical BER analysis is reasonably consistent with simulated one and has relatively low computational complexity as compared to the PEXIT algorithm. Consequently, the proposed BER analysis can be used to accurately evaluate the error performance and predict the decoding threshold for the STBC-JCNC systems.

\section{Acknowledgements}
This work was supported by the NSF of China (nos. 61271241 and 61001073), the European Union-FP7 (CoNHealth, no. 294923), and the
Fundamental Research Funds for the Central Universities (No. 201112G017).

\begin{thebibliography}{10}
\providecommand{\url}[1]{#1}
\csname url@samestyle\endcsname
\providecommand{\newblock}{\relax}
\providecommand{\bibinfo}[2]{#2}
\providecommand{\BIBentrySTDinterwordspacing}{\spaceskip=0pt\relax}
\providecommand{\BIBentryALTinterwordstretchfactor}{4}
\providecommand{\BIBentryALTinterwordspacing}{\spaceskip=\fontdimen2\font plus
\BIBentryALTinterwordstretchfactor\fontdimen3\font minus
  \fontdimen4\font\relax}
\providecommand{\BIBforeignlanguage}[2]{{%
\expandafter\ifx\csname l@#1\endcsname\relax
\typeout{** WARNING: IEEEtran.bst: No hyphenation pattern has been}%
\typeout{** loaded for the language `#1'. Using the pattern for}%
\typeout{** the default language instead.}%
\else
\language=\csname l@#1\endcsname
\fi
#2}}
\providecommand{\BIBdecl}{\relax}
\BIBdecl

\bibitem{Zhang2006}
S.~Zhang, S.-C. Liew, and P.~Lam, ``{Hot topic: physical layer network
  coding},'' in \emph{Proc. MobiCom'06}, Sept. 2006, pp. 358--365.

\bibitem{6133672}
H.~Gao, X.~Su, T.~Lv, T.~Wang, and Z.~Wang, ``{Physical-layer network coding
  aided two-way relay for transmitted-reference UWB networks},'' in \emph{Proc.
  IEEE Global Telecommun. Conf. (GLOBECOM)}, Dec. 2011, pp. 1--6.

\bibitem{6364770}
L.~K.~S. Jayasinghe, N.~Rajatheva, and M.~Latva-aho, ``{Joint pre-coder and
  decoder design for physical layer network coding based MIMO two-way relay
  system},'' in \emph{Proc. IEEE Int. Conf. Commun. (ICC)}, Jun. 2012, pp.
  5645--5649.

\bibitem{5545660}
D.~To, J.~Choi, and I.-M. Kim, ``{Error Probability analysis of bidirectional
  relay systems using Alamouti scheme},'' \emph{IEEE Commun. Lett.}, vol.~14,
  no.~8, pp. 758--760, Aug. 2010.

\bibitem{6380933}
P.~Chen, L.~Wang, and J.~He, ``{Physical-layer network coding and precoding for
  end nodes using Alamouti scheme},'' in \emph{Proc. Int. Symp. Commun. Inf.
  Technol. (ISCIT)}, Oct. 2012, pp. 417--422.

\bibitem{5072363}
S.~Zhang and S.-C. Liew, ``{Channel coding and decoding in a relay system
  operated with physical-layer network coding},'' \emph{IEEE J. Sel. Areas
  Commun.}, vol.~27, no.~5, pp. 788--796, Oct. 2009.

\bibitem{5493997}
Y.~Lang and D.~Wubben, ``{Generalized joint channel coding and physical network
  coding for two-way relay systems},'' in \emph{Proc. IEEE Veh. Technol. Conf.
  (VTC)}, May 2010, pp. 1--5.

\bibitem{5683819}
D.~Wubben and Y.~Lang, ``{Generalized sum-product algorithm for joint channel
  decoding and physical-layer network coding in two-way relay systems},'' in
  \emph{Proc. IEEE Global Telecommun. Conf. (GLOBECOM)}, Dec. 2010, pp. 1--5.

\bibitem{6381454}
P.~Chen, Y.~Fang, L.~Wang, and F.~C.~M. Lau, ``{Decoding generalized joint
  channel coding and physical network coding in the LLR domain},'' \emph{IEEE
  Signal Process. Lett.}, vol.~20, no.~2, pp. 121--124, Feb. 2013.

\bibitem{6125282}
T.~Huang, T.~Yang, J.~Yuan, and I.~Land, ``{Convergence analysis for
  channel-coded physical layer network coding in Gaussian two-way relay
  channels},'' in \emph{Proc. Int. Symp. Wireless Commun. Syst. (ISWCS)}, Nov.
  2011, pp. 849--853.

\bibitem{380145}
T.~Eng and L.~Milstein, ``{Coherent DS-CDMA performance in Nakagami multipath
  fading},'' \emph{IEEE Trans. Commun.}, vol.~43, no. 234, pp. 1134--1143, Feb.
  1995.

\bibitem{Molisch04ieee802.15.4a}
A.~Molisch \emph{et~al}., \emph{{IEEE 802.15.4a channel model-final
  report}}.\hskip 0.2em plus 0.1em minus 0.4em\relax IEEE 802.15-04-0662-00-004a,
  2004.

\bibitem{Fang2012}
Y.~Fang, P.~Chen, L.~Wang, F.~C.~M. Lau, and K.~K. Wong, ``{Performance
  analysis of protograph-based LDPC codes with spatial diversity},'' \emph{IET
  Commun.}, vol.~6, no.~17, pp. 2941--2948, Dec. 2012.

\bibitem{730453}
S.~Alamouti, ``{A simple transmit diversity technique for wireless
  communications},'' \emph{IEEE J. Sel. Areas Commun.}, vol.~16, no.~8, pp.
  1451--1458, Oct. 1998.

\bibitem{4155107}
A.~Abbasfar, D.~Divsalar, and K.~Yao, ``{Accumulate-repeat-accumulate codes},''
  \emph{IEEE Trans. Commun.}, vol.~55, no.~4, pp. 692--702, Apr. 2007.

\bibitem{485714}
J.~Hagenauer, E.~Offer, and L.~Papke, ``{Iterative decoding of binary block and
  convolutional codes},'' \emph{IEEE Trans. Inf. Theory}, vol.~42, no.~2, pp.
  429--445, Feb. 1996.

\bibitem{Fang2013}
\BIBentryALTinterwordspacing
Y.~Fang, K.~K. Wong, L.~Wang, and K. F. Tong, ``{Performance analysis of protograph LDPC codes for Nakagami-$m$ fading relay channels},''
  \emph{Accepted by IET Commun.}, Apr. 2013. [Online]. Available:
  \url{http://arxiv.org/abs/1304.6614}
\BIBentrySTDinterwordspacing

\bibitem{6253209}
Y.~Fang, P.~Chen, L.~Wang, and F.~C.~M. Lau, ``{Design of protograph LDPC codes
  for partial response channels},'' \emph{IEEE Trans. Commun.}, vol.~60,
  no.~10, pp. 2809--2819, Oct. 2012.

\bibitem{5751586}
B.~S. Tan, K.~H. Li, and K.~Teh, ``{Performance analysis of LDPC codes with
  maximum-ratio combining cascaded with selection combining over Nakagami-$m$
  fading},'' \emph{IEEE Trans. Wireless Commun.}, vol.~10, no.~6, pp.
  1886--1894, Oct. 2011.

\bibitem{957394}
S.~ten Brink, ``{Convergence behavior of iteratively decoded parallel
  concatenated codes},'' \emph{IEEE Trans. Commun.}, vol.~49, no.~10, pp.
  1727--1737, Oct. 2001.

\end{thebibliography}
\end{document}